          [2001/04/25 1.1 (PWD)]
\documentclass[a4paper,twocolumn]{esapub} 
\usepackage{natbib}
\usepackage{graphicx}

\title{Sungrazing comets as source of pickup ions at Earth orbit and Ulysses}
\author{Maciej Bzowski}
\author{Ma\l gorzata Kr\'olikowska }
\affil{Space Research Centre PAS, Bartycka 18A, 00-716 Warsaw,
Poland }

\begin{document}


\maketitle

\begin{abstract}
Most of the sungrazing comets observed by LASCO at SOHO belong to
the Kreutz group of comets and follow trajectories that are
tightly clumped in space. Statistical analysis of 9~years of SOHO
observations suggests that the true apparition rate of these
comets is as high as one every other day. Practically all these
comets break up before perihelion passage. Their material is
dissociated and ionized, and subsequently transported away from
the Sun as pickup ions in the solar wind. Their mean mass flux is
about $3.1\cdot 10^{4}$~g$\cdot$s$^{-1}$. Since the breakup occurs
between 40 and 4~solar radii and the ionization is almost
immediate, the expected location of these ions in the phase space
is close to the location of the inner source of pickup ions.
Assuming radial propagation, the cometary pickups should be
observable at Earth between August and January, with peak
probability at the end of September. At Ulysses, they should be
observable approximately between $-$25 and 40~degrees ecliptic
latitude during the fast latitude scans, the first of which
occurred in 1995 and the second in 2001. The population of
cometary pickup ions should be augmented by about 40\% by solar
wind protons as a result of charge exchange with the cometary
neutral hydrogen and oxygen atoms and subsequent reionization of
the newly-created Energetic Neutral Atoms, streaming with respect
to the solar wind. In total, the average flux of the pickup ions
related to the sungrazing comets at 1~AU should be about $1.6\cdot
10^{5}$~g$\cdot$s$^{-1}\cdot$sr$^{-1}$ within the detection area
(and null outside it). This value is comparable to the flux of the
inner source-pickup ions.

\end{abstract}

\section{Introduction}
Pickup ions (PUIs) in the solar wind are former neutral atoms of thermal
energy, ionized and picked up by the magnetic field frozen in the solar
wind \citep{fahr:73}. The main source of PUIs is
neutral interstellar gas. PUIs from interstellar helium were discovered
by \citet{mobius_etal:85a} and from interstellar hydrogen by
\citet{gloeckler_etal:93a}. The inner source of PUIs was discovered by
\citet{geiss_etal:95a}. Observations of the PUI distribution function
\citep{gloeckler_geiss:98a} and modeling suggest that the peak source
rates of these ions occur between 10 and $30~R_{S}$ and its elemental
composition is similar to the composition of solar wind
\citep{gloeckler_etal:00a}.

Several explanations of the PUI inner source were proposed, all of them
involving some kind of interaction of interplanetary dust with the solar
wind: recycling of the solar wind ions on dust grains
\citep{gloeckler_etal:00a, schwadron_etal:00}, neutralization of solar
wind ions on nanometer-sized dust grains
\citep{wimmer-schweingruber_bochsler:03}, and ionization and pickup of
products of collisions of dust grains close to the Sun
\citep{mann_czechowski:05a}. None of these scenarios is able to fully
explain the observations and one of the important problems is the lack
of sufficient amount of dust so close to the Sun
\citep{mann_czechowski:05a}. A more thorough discussion of these scenarios
is offered by \citet{allegrini_etal:05a}.

It seems that there might be more than just one ``inner source'' of
PUIs. In a recent paper \citet{bzowski_krolikowska:05a} (further on: BK05)
suggested that at least a part of the inner source of pickup ions
might be the material released by sungrazing comets from the Kreutz
group. In this short communications we recapitulate and expand
these ideas and results.

\section{Kreutz sungrazers as source of pickup ions}

Shortly after beginning of operations, the LASCO coronograph onboard the
SOHO spacecraft discovered a stream of comets approaching the Sun to a
few solar radii \citep{biesecker_etal:02}. Observations show that 85\%
of these comets belong to a single group of comets, discovered by Kreutz
\citep{kreutz:1901} and carrying its name. The Kreutz comets seem to
have been present since their discovery until now, as evidenced
by limited optical and space-borne observations listed by BK05; they
were present at the discovery of the inner source by Ulysses and have
been observed until now. Based on the yearly pattern of perihelion times
rate observed during first 2 years of LASCO operations, which features
peaks in June and December, \citet{biesecker_etal:02} concluded that the
true Kreutz rate is much higher than observed due a geometrical
instrument selection effect and equal to about 1 every other day. This
rate was confirmed by BK05 based on a 9 year's statistics.

\begin{figure}
\centering
\includegraphics[width=1.0\linewidth]{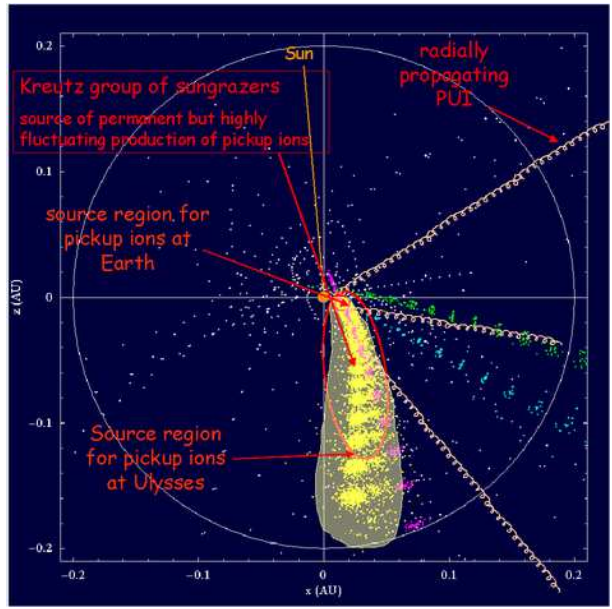}
\caption{Cartoon figure illustrating the source region and observing
conditions of the PUIs from the sungrazing comets. Projected on the x-z
plane of the heliocentric ecliptic system (where x-axis points to the
vernal equinox point) is the region occupied by trajectories of the
Kreutz comets (yellow dots in the plume-like contour). The Sun (drawn
in-scale) is at the center. Also shown are projections of trajectories
of the Marsden (cyan), Meyer (magenta), and Kracht (green) groups of
comets and of the observed ``stragglers'' (white). The region occupied
by the Kreutz group is the source region of the PUIs from the material
released by the comets discussed in the paper. Assuming radial
propagation of PUIs from the birth place (illustrated by orange 
spirals), Ulysses and Earth-bound spacecraft are able to see the cometary
PUIs when their orbits intersect with the radial projection of the
source regions. 
\label{fig:cartoon}}
\end{figure}

The Kreutz comets follow very similar, highly-elliptical (almost
parabolic) orbits and disintegrate between 40 and $4~R_{S}$ in a
well-constrained region of space. Based on analysis by
\citet{sekanina:03}, BK05 estimated the mass influx from
the Keutz comets at $3.1 \cdot 10^4$~g~s$^{-1}$. As close to the Sun as
the Kreutz comets disintegrate, the whole material is very quickly (minutes
to hours) dissolved into atoms and simple molecules, ionized and
picked up by the solar wind. The ionization processes are charge
exchange between the cometary neutral atoms and solar wind protons,
impact of solar wind electrons and ionization by solar EUV photons. The
charge exchange effectively contributes an extra source of the
comets-related PUIs: one of the reaction products is an Energetic
Neutral H-atom (H-ENA) which, being relatively slow close to the Sun,
eventually gets ionized. Since before ionization it had a non-zero velocity with
respect to the accelerating solar wind in the location of its
reionization, it becomes and extra PUI proton. A more thorough discussion
of this reaction channel and related physics is provided by BK05. They
estimate that the H ENA reionization channel increases the number of the
comet-related PUI's by about 40\% and that the cometary PUI composition
(by number) will be 43\% of H, 27\% of O, 25\% of C, and 5\% of other
elements, as Mg, Si, Fe etc. It is worth noticing as well that the
cometary dust, especially the grains released before breakup, are
subject to the same interaction with solar wind as in all proposed
``dusty'' scenarios of the inner source origin, and as such are able to
produce, e.g., neon PUI, also observed in the inner source, and makes extra
targers for the dust-related mechanisms of inner-source PUI production.
Since, however, details of the interaction of dust grains with solar wind
during such a short time as it remains between grain release at the breakup
of nucleus and evaporation of these grains in the heat from the Sun,
it is difficult to assess qualitatively the extra intensity of inner source
from this channel.

Assuming radial propagation of PUIs from the pickup location away from
the Sun, as schematically illustrated in Fig.~\ref{fig:cartoon}, BK05
predicted that an Earth-bound spacecraft should be in a favorable
position to observe the cometary PUIs since the end of July until
beginning of November each year, with the peak probability of detection
at the end of September, and Ulysses on its polar orbit around the Sun
should be inside the detection region between Feb.10, 1995 and May 14,
1995, and again during the next revolution, between April 23 and July
23, 2001, while in the ecliptic latitude band between $-26^{\circ}$ and 
$41^{\circ}$. 
\begin{figure}
\centering
\includegraphics[width=1.0\linewidth]{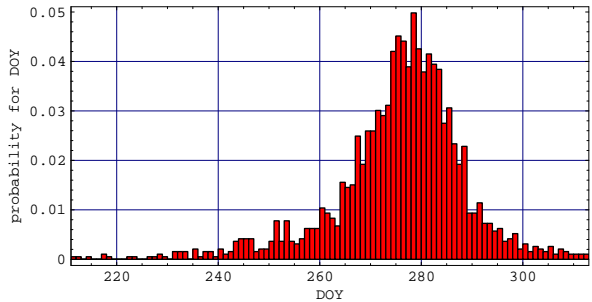}
\caption{Probability distribution function (normalized to 1)
for detection time of cometary PUIs from Earth-bound spacecraft.
Estimate based on orbits of all observed Kreutz comets.
\label{fig:earthHisto}}
\end{figure}

\begin{figure}
\centering
\includegraphics[width=1.0\linewidth]{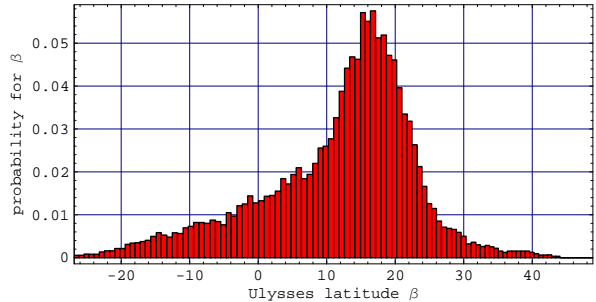}
\caption{Probability distribution function (normalized to 1) for 
ecliptic latitude for detection of comatary PUIs from Ulysses 
in Spring of 1995 and 2001. Estimate based on orbits of all 
observed Kreutz comets. \label{fig:ulyHisto}}
\end{figure}

In this paper we refine this prediction, as illustrated in 
Fig.~\ref{fig:earthHisto} for Earth-bound spacecraft and in 
Fig.~\ref{fig:ulyHisto} for Ulysses. BK05 assessed the average count 
rate at 1~AU at 12.5~s$^{-1}$~cm$^{-2}$. Fig.~\ref{fig:earthHisto} shows
that the probability of detection is highly peaked within the indicated
time interval of DOY 215 -- DOY 310 and that the core of the probability
distribution function is contained between DOY 260 and DOY 290, 
which corresponds to September 17 and October 17. The mean count rate during 
this interval should be $\sim 3$-fold higher than originally estimated, reaching
an easily-detectable value of $\sim 40$~s$^{-1}$~cm$^{-2}$. Ulysses
should see the core of the probability distribution function when 
between $5^{\circ}$ and $25^{\circ}$ north ecliptic latitude in Spring of
1995 and 2001.

\section{Discussion}
BK05 considered a hypothesis that the Kreutz sungrazers could be in fact the sole
inner source of PUIs. If this should be the case (and if the hypothesis of 
radial propagation of the cometary PUIs from their birth place holds), then
the inner source should be observable only during the intervals indicated by BK05.
\citet{allegrini_etal:05a} checked this hypothesis, searching for 
the inner source PUIs in Ulysses data during three selected intervals
outside the intervals suggested by BK05, and found that the inner source
is there, correlated with the solar wind flux and stable over the solar cycle. 
Hence the sungrazing comets cannot be the sole inner source of PUI. Nevertheless,
analysis of inner source observations during the intervals suggested by 
BK05 and in this paper still needs to be done. \citet{allegrini_etal:05a}
report lower production of C and O (5 and 7 x $10^5$~g~s$^{-1}$, respectively) 
than inferred by \citet{geiss_etal:96a} ($2\cdot10^6$~g~s$^{-1}$ each), 
who refer to Ulysses observations performed in the latitude band including 
the band suggested by BK05 as prone for detection of cometary PUIs. 

The indication that actually observed inner-source PUIs are partly due
to the Kreutz comets would be a reduction of Ne content in the inner source
PUI flux during the suggested
intervals with an increase of the net inner source PUI flux. Apart from
these, the cometary PUIs are hardly discernible from the ``regular''
inner source PUIs. As pointed out both by BK05 and
\citet{allegrini_etal:05a}, they should occupy the same location in the
phase space. Further, the composition should be similar, though the Ne
content should be reduced since the only way to obtain Ne PUIs from the
cometary material would be to neutralize the solar wind Ne ions on the
cometary dust grains, which have short lifetimes at breakup. Since the
main ionization channels for cometary material are charge exchange with
solar wind protons and electron impact, the cometary PUI flux should be
correlated with the solar wind flux similarly as observed in the case of
regular inner source PUIs. Finally, contrary to the suggestion by
\citet{allegrini_etal:05a}, counts from the cometary PUIs should show
statistcs relevant for a randomly-distributed source inside a
well-constrained region of space. The apparition rate of the
Kreutz comets (1 every two days) is well comparable with the travel time
of PUIs from the breakup region to 1~AU and Ulysses and the apparition
times of the Kreutz sungrazers follow Poisson distribution. Therefore also
the PUI parcels from sungrazing comets should show a Poisson statistcs,
though their source region will be constrained in space.

Summarizing: on one hand the intensity of inner source is still
uncertain due to strong fluctuations and low count rate \citep{allegrini_etal:05a}.
On the other hand, an independent estimate of mass inflow from sungrazing
comets is available and this entire mass is converted into PUI with
the inner source characteristics. Hence comparing observations of the inner s
ource PUIs inside and outside
the regions suggested by BK05 and in this paper could bring an important
clue on the true rate of the inner source and consequently provide evidence on its
nature.

\section*{Acknowledgments}
This research was supported by the Polish State Committee for Scientific Research Grant
1P03D~009~27.
\bibliographystyle{aa}
\bibliography{iplbib}

\end{document}